\documentclass[10pt, twocolumn]{IEEEtran_v15}

\usepackage{graphicx}

\newcommand{\ig}{\includegraphics*[width=3.5in]}

\begin{document}

% now set the left margin to be about 0.65in
% we don't use the typically given 0.7in
% because we want to preserve the 21pica
% text columns (with 1pica columnsep)
% AND maintain symmetric left and right margins
% on letter paper
\setlength{\evensidemargin}{-0.35in}
\setlength{\oddsidemargin}{-0.35in}
 
% choose a text height which gives 1 inch
% top and bottom margins. Text height is
% normally 58pica (9.6306in) in Transactions.
% we'll use 9in here
\setlength{\textheight}{9in}
 
% LaTeX assumes a 1 inch margin offset
% so if we zero every header at the top,
% we'll be left with a 1 inch top margin
\setlength{\topmargin}{0in}
\setlength{\headheight}{0in}
\setlength{\headsep}{0in}
 
% Turn on center justification of the short figure captions.
% This command requires IEEEtran.cls V1.4 or later!
\centerfigcaptionstrue

\title{\Large\bfseries High-Frequency Acoustic Sediment Classification in Shallow Water} 

% IEEEtran normally already uses \large (12pt) for author's names.
% However, we must add \normalsize for the affiliation and use line
% separators as desired.
% Note: V1.5 and later support the use of the \and operator to list multiple
% authors of different affiliations. Within the IEEEtran.cls file, this is
% done by enclosing the \author{} text within a tabular environment - just
% like article.cls does. A consequence of this is that you now must
% repeat
% font size commands for each line and you cannot enclose multiple lines
% within an environment.
% i.e., you CANNOT do: \author{\large name \\ {\normalsize line2 \\ line3} }
% (at least this is true on my TeTex system - even under article.cls)
%

\author{\large Frank W. Bentrem{\normalsize\textsuperscript{*}}, John
Sample{\normalsize\textsuperscript{*}}, Maria
T. Kalcic{\normalsize\textsuperscript{*}}, and Michael~E.
Duncan{\normalsize\textsuperscript{\dag}}\vspace{1ex}\\
\normalsize \textsuperscript{*}Marine Geosciences Division, Naval Research Laboratory, Stennis Space Center, Mississippi 39529\\
\normalsize\textsuperscript{\dag}Planning Systems Incorporated, Slidell,
Louisiana 70458-1350.}

\maketitle
\thispagestyle{plain}
\pagestyle{plain}

\setcounter{page}{7}

\begin{abstract}
A geoacoustic inversion technique for high-frequency (12 kHz) multibeam
sonar data is 
presented as a means to classify the seafloor sediment in shallow
water (40--300 m). 
The inversion makes use of backscattered data at a variety of grazing 
angles to estimate mean grain size. The need for sediment type and the 
large amounts of
multibeam data being collected with the Naval Oceanographic Office's Simrad EM 121A systems, have 
fostered the
development of algorithms to process the EM 121A acoustic backscatter into 
maps of sediment type. The
APL-UW (Applied Physics Laboratory at the University of
Washington) backscattering model is used with simulated annealing
to invert for six geoacoustic parameters. For the inversion, three of the 
parameters 
are constrained according to empirical correlations with mean 
grain size, which is introduced as an unconstrained parameter. The four 
unconstrained (free) parameters are mean 
grain size, sediment volume interaction, and two seafloor roughness 
parameters. Acoustic sediment
classification is performed in the Onslow Bay region off the coast of
North Carolina using data from the 12kHz Simrad EM 121A multibeam sonar 
system. Raw hydrophone data is beamformed into 122 beams with a 120-degree 
swath on the ocean floor, and backscattering strengths are calculated for 
each beam and for each ping. Ground truth consists of 68 grab 
samples in the 
immediate vicinity of the sonar survey, which have been analyzed for mean 
grain size. Mean grain size from the inversion shows 90\% agreement with the 
ground truth and may be a useful tool for high-frequency acoustic sediment 
classification in shallow water.

\end{abstract}

\section{Introduction}
The U.\ S. Navy has great interest in seafloor characterization due to its
importance in shallow-water operations, such as landing
operations, mine burial, and safety of navigation. Determining a suitable route for communications cables, requires
detailed knowledge of the seafloor and is another application for
characterization of the ocean bottom.

Obtaining and analyzing physical core samples or grab samples provides
an accurate characterization of the seafloor, however, it is a
time-consuming process and is not generally performed with sufficient
coverage on an ocean survey. As an alternative, acoustic seafloor
characterization allows adequate coverage in much less time and, since
sonar data is often collected on surveys, no additional data
collection is required. The acoustic data evaluated in this
paper was collected in Onslow Bay with the 12 kHz Simrad EM 121A Multibeam Echo Sounder.

\subsection{Sediment Types}

One of the most useful descriptors for bottom characterization is
sediment type based on the mean grain diameter, which can range from
clay ($\approx$ 0.0039 mm) to boulders ($\approx$ 256 mm) or greater. A phi value
$\phi$ scale conveniently represents the mean grain size according to

\begin{equation}
\phi=-\log_2\frac{d}{d_0},
\label{eq:phi}
\end{equation}

\noindent
where $d$ is the mean grain diameter in mm and $d_0$ is the reference
diameter 1 mm.
Approximate $\phi$ values for selected sediments
are given in Table~\ref{tab:sediment} according to the Wentworth scale
\cite{wentworth22}.

\begin{table}

\caption{Sediment Types}

\label{tab:sediment}

\begin{center}

\begin{tabular}{ccc}

\\[-1cm]

\hline

\\[-.35cm]

Phi Value $\phi$ & Mean Grain Diameter & Sediment Type\\

& (mm) & \\

\\[-.35cm]

\hline

\\[-.35cm]

$\le$ ($-1.0$) & $\ge 2.0$ & gravel/rock\\

\\[-.35cm]

($-1.0$) -- $4.0$ & $0.06$ -- $2.0$ & sand\\

\\[-.35cm]

$4.0$ -- $8.0$ & $0.004$ -- $0.06$ & silt\\

\\[-.35cm]

$> 8.0$ & $< 0.004$ & clay\\

\\[-.35cm]

\hline

\\[-.35cm]

\end{tabular}

\end{center}

\end{table}

\subsection{Onslow Bay}

Onslow Bay off the coast of North
Carolina is a challenging region for high-frequency acoustic sediment classification because the bottom is dynamic (sediment drift) \cite{riggs98}, heterogeneous
in areas \cite{cleary68,dealteris96} with shells, etc., mixed with the sediment, and is often
composed of a hard bottom \cite{dealteris96,riggs96} covered with only a thin (few centimeters or less) layer of sediment. 
The sonar data set from this survey is raw hydrophone data along the three
parallel shiptracks depicted in Fig.~\ref{fig:onslow} that are more or less parallel to the
coastline. Shiptrack 1 has 250 -- 300 m water depth and is farthest
from shore near the continental-shelf break. The seafloor slopes up to
0.5$^\circ$ in a direction perpendicular to the ship's
heading. Shiptracks 2 and 3 are in shallower water (40 -- 60 m) about
80km from shore, seafloor here on the shelf is relatively flat.

\begin{figure}
 
\centerline{\ig{shiptrack.eps}}
 
\caption{Shiptracks and collocated grab samples in the Onslow Bay area.}
 
\label{fig:onslow}
 
\end{figure}

\subsection{Hydrophone Data Processing}

Each port and starboard arrays are comprised of 64
hydrophones.  
Each array is steered
between -60 and 60 degrees (negative being in the port direction) in
one-degree increments. However, both port and starboard arrays are steered at
0 degrees so there are 122 steer directions (beams).

\subsubsection{Preprocessing}
 
For each ping, header, raw data, and PAM (Power Amplifier Monitor)
records are read from the tape or a file. A number of samples (usually
1028) from each ping are taken from
the raw data record.  Where applicable the beginning sample is selected
according to the value of the Time-Varying Gain TVG and a hard-coded
threshold. The effects of Programmable Gain PG, Fixed Gain FG, and
Time-Varying Gain TVG are removed from the data. These computations
are made in linear space based on values obtained from the header
record. The data, now in units of
digital number DN, are converted to sound pressure level
SPL.  Values for this conversion are taken from the header record.
 
At this point the data are still at baseband. To
beamform, the data are shifted to the original center frequency
(12 kHz).  To avoid aliasing the basebanded
data must be resampled at a higher rate than the original sampling
rate of 2.5 kHz.  The resample factor used is 16, so the resampling
rate is 40 kHz.  The interpolation is done via a Fast
Fourier Transform (FFT).  The
slow data are transformed to the frequency domain with a large FFT,
shifted, then transformed back to the time domain.
 
After the shift to 12 kHz, the average roll, pitch, heave and yaw
for the given ping are computed.  These values are then used to adjust
the absolute locations (in software) of the receiver staves in the
array and will enable
the beams to be steered to consistent beam angles relative to the seafloor.
 
Following the motion correction the data are beamformed by phase adjusting the frequency
domain data according to the receiver locations and desired
steering angles.  Taking the inverse FFT of these data yields a sound pressure
$P$ time series for each
steering angle. The travel time of the bottom return is identified for each angle and an
acoustic ray is traced out (here a constant sound speed
profile is used because of a negligible sound speed gradient) to the
corresponding bottom returns in order to obtain grazing angle.  

The sound
pressure for the $j$th time sample of the $i$th beam is denoted
$P_{ij}$. The data are converted to dB re $\mu$Pa and, based on the known
geometry, the sonar equation is solved for bottom backscatter.

\subsubsection{Backscattering Strength}

Backscattering strength $BS$ is defined as 

\begin{equation}
BS=10\log_{10}\frac{I_b}{I_{inc}},
\label{eq:backsc}
\end{equation}

\noindent
where $I_b$ is the backscattered sound intensity from an area of 1~m$^2$ and
$I_{inc}$ is the incident intensity at 1~m from the source \cite{urick83}. The backscattering
strength can be determined from the data by using the sonar equation

\begin{equation}
BS=RL-SL+2TL-IA,
\label{eq:sonar}
\end{equation}

\noindent
where $RL$ is the reverberation level (from the beamformed time
series), $SL$ is the source level, $TL$ is the transmission loss in dB, and $IA$ is the insonified area in dB re m$^2$. The insonified area is the area contributing to the
received intensity and is computed using the 3 dB beam footprint,

\begin{equation}
IA=10\log_{10}\biggl\{2R^2\sin\frac{\theta_t}{2}\biggl[\cot\biggl(\theta+\frac{\theta_r}{2}\biggr)-\cot\biggl(\theta-\frac{\theta_r}{2}\biggr)\biggr]\sin\theta\biggr\},
\label{eq:footprint}
\end{equation}

\noindent
or using the pulse length,

\begin{equation}
IA=10\log_{10}\biggl(\frac{c\tau R\sin\frac{\theta_t}{2}}{\cos\theta}\biggr),
\label{eq:pulse}
\end{equation}

\noindent
whichever is smaller, where $R$ is the slant range to the bottom,
$\theta_t$ is the transmit beam width, $\theta_r$ is the receive beam
width, $c$ is the water sound speed in m/s, and $\tau$ is the pulse
duration in s. The insonified
area for several pressure time samples $P_{ij}$ normally fall within the beam footprint,
and the reverberation level for the $i$th beam $RL_i$ is averaged over these time samples,

\begin{equation}
RL_i=10\log_{10}\biggl(\frac{\sum_{j_0}^{j_1}P_{ij}^2}{j_1-j_0+1}\biggr),
\label{eq:reverb}
\end{equation}

\noindent
where $j_0$ and $j_1$ are the first and last time samples whose
insonified areas lie within the $i$th beam's footprint.

\begin{table*}

\caption{Model Input Parameters}

\label{tab:input}

\begin{center}

\begin{tabular}{ccc}

\\[-1cm]

\hline

\\[-.35cm]

Parameter & Symbol & Description\\

\hline

\\[-.35cm]

Density Ratio & $\rho$ & $\frac{\hbox{density in
sediment}}{\hbox{density in water}}$\\

\\[-.2cm]

Sound Speed Ratio & $\nu$ & $\frac{\hbox{sound speed in
sediment}}{c}$\\

\\[-.2cm]

Loss Parameter & $\delta$ & $\frac{\hbox{imaginary wavenumber in
sediment}}{\hbox{real wavenumber}}$\\

\\[-.2cm]

Spectral Strength & $\beta$ & Bottom height spectrum strength\\

\\[-.35cm]

Spectral Exponent & $\gamma$ & Bottom height spectrum exponent\\

\\[-.2cm]

Volume Parameter & $\sigma$ & $\frac{\sigma_v}{\hbox{sediment attenuation coefficient}}$\\

\\[-.35cm]

\hline

\\[-.35cm]

\end{tabular}

\end{center}

\end{table*}

\section{Backscatter Model}

The APL-UW backscatter model presented by Mourad and Jackson
\cite{mourad89,APL94} treats the seafloor as a statistically homogeneous
fluid and predicts backscattering strength $BS$ as a function of
grazing angle $\theta$. The roughness of the bottom is described in
this model by
the bottom height spectrum.

\begin{equation}
W=\beta\biggl(\frac{2\pi fh}{c}\biggr)^{-\gamma},
\label{eq:rough}
\end{equation}

\noindent
where $h$ is the reference height 1cm. The Mourad-Jackson model is
valid for all frequencies between 10 and 100 kHz and is used here to
represent the acoustic backscatter from the seafloor.

Table~\ref{tab:input} lists the six model input parameters, which,
along with the sonar frequency $f$ and sound speed $c$ in water at the
seafloor, determine both the roughness backscattering cross section
$\sigma_r(\theta)$ and volume backscattering cross section
$\sigma_v(\theta)$. The six input parameters are dimensionless except
for $\beta$ which has units of cm$^4$. Combining these backscatter contributions from roughness
(acoustic reflections from a randomly rough surface) and volume
interaction (scattering of penetrating sound from sediment
inhomogeneities) results in

\begin{equation}
BS(\theta)=10\log_{10}(\sigma_r+\sigma_v).
\label{eq:backsc_calc}
\end{equation}

\section{Data Inversion}

The inversion problem is finding the set of input parameters that best
fits the given data set. That is, which set of parameters minimizes
the difference between the $BS$ vs. $\theta$ curve and the measured
backscatter data. The sum of the squares of the data deviations
from the model prediction is used as the measure for goodness of fit.

\subsection{Parameter Constraints}

If the six input parameters are unconstrained, the parameter space to be
searched is six-dimensional. However, since correlations exist among
some of the
parameters, many solution parameter sets represent
solutions that are physically unlikely. Hamilton and Bachmann \cite{hamilton72,hamilton82} describe a relationship
between the parameters $\rho$ and $\nu$ and relate both to the mean grain
size ($\phi$) of the seafloor sediments. Mourad and Jackson
\cite{mourad89} parameterize $\rho$, $\nu$, and $\delta$ according to
$\phi$ values emphasizing the top few tens of centimeters of
sediment, and the parameterization has been generalized to include
coarse sand \cite{APL94}. (Some correlation exists between $\delta$ and $\phi$, and
the effect of physically meaningful values of $\delta$ on the
$BS$ vs. $\theta$ curve is negligible.) Gott \cite{gott95} has
used the idea of
constraining some of the model parameters with some success. In
addition the parameters used should be restricted to values that are
physically likely. The parameter ranges used here are presented in Table~\ref{tab:range}.

\begin{table}

\caption{Parameters Ranges}

\label{tab:range}

\begin{center}

\begin{tabular}{cc}

\\[-1cm]

\hline

Parameter & Range\\

\hline

$\phi$ & ($-1.0$) -- $9.0$\\

\\ [-.35cm]

$\beta$ & within factor of 2 of APL-UW\\

& parameterization \cite{APL94}\\

$\gamma$ & $2.4$ -- $3.9$\\

$\sigma$ & $0.00$ -- $0.02$\\

\hline

\\[-.35cm]

\end{tabular}

\end{center}

\end{table}

The parameter space to be searched is now 4-dimensional ($\phi$,
$\beta$, $\gamma$, $\alpha$), and, since the backscatter model is
highly nonlinear, one must be careful not to simply find one of the
many local solutions. Two of the most common global search methods are
simulated annealing and genetic algorithms. Both are suitable for most
nonlinear problems. Simulated annealing (SA) is the
best-fit search routine used here (e.g. see \cite{press92}).

\subsection{Simulated Annealing}

With the SA approach one searches the
parameter space by continuously stepping to a new point in parameter space
and computing the sum of the squares $E$ for the data
point residuals. $E$ is also known as the cost function.  If the cost decreases from the previous location, the step is
accepted.  If, however, the cost increases, the step is only
occasionally accepted.  The probability $p$ that a higher-cost step is accepted
depends both on the amount of increased cost $\Delta E$ and on a variable referred to as
temperature $t$ according to the Boltzmann distribution,

\begin{equation}
p=e^{-\Delta E / t}.
\label{eq:prob}
\end{equation}

\noindent
This process is known as the Metropolis algorithm
\cite{metropolis53}. Local minima are escaped because of the steps of
increased cost.  The temperature variable is gradually decreased until the
probability of a higher-cost step is zero.  The stepsize is also
reduced slowly as the algorithm settles into
the global minimum. 

\section{Results}

\subsection{Slope Region}

The data for shiptrack 1 (farthest from shore) was grouped into bins
of 200 pings covering an area of seafloor approximately 3 km$\times$1 km. The
backscattering strengths in each bin were averaged according to grazing angle,
and a best-fit parameter set was found via simulated annealing for the averaged
data. To illustrate Fig.\ \ref{fig:best_fit} shows backscatter data for the first
200 pings for shiptrack 1 along with the SA best-fit model curve.
A
comparison of inversion phi values with the analyzed grab samples
is shown in Fig.~\ref{fig:tape1}.
All 58 inversions for the 200-ping bins result in phi values
indicating medium or fine grades of sand. The inversion phi values are
in most cases only slightly greater than medium sand measured at the
nearest grab sample location.  

\begin{figure}
 
\centerline{\ig{best_fit.eps}}
 
\caption{Backscatter data from the first 200 pings from shiptrack 1 and
the best-fit model curve. The inversion indicates $\phi=1.41$, while the
nearest grab sample shows $\phi=2.08$.}
 
\label{fig:best_fit}
 
\end{figure}

\begin{figure}
 
\centerline{\ig{tape1.eps}}
 
\caption{Comparison of phi values from inversion with the measured grab samples in
the slope region.}
 
\label{fig:tape1}
 
\end{figure}  

\subsection{Shelf Region}

Because of a higher ping rate, the backscatter data from shiptracks 2 and 3
are binned in groups of 500 pings each.
Figures~\ref{fig:tape2} and \ref{fig:tape3} (closest to shore) compare
the phi values from the SA inversion to the grab samples in the shallower
water ($\approx$ 40--60 m) with 61 of the 71 inversion phi values
matching the nearest grab sample sediment type. Forty-three grab samples in
the shelf region show a medium
or coarse sand bottom with one sample indicating gravel. The
inversion yields 60 sand, 9 gravel, and 2 clay values. This region
exhibits greater variation in phi values than the slope region.

\begin{figure}
 
\centerline{\ig{tape2.eps}}
 
\caption{Comparison of phi value from inversion with grab sample analysis in
the shelf region along shiptrack 2.}
 
\label{fig:tape2}
 
\end{figure}

\begin{figure}
 
\centerline{\ig{tape3.eps}}
 
\caption{Comparison of phi value from inversion with grab sample analysis in
the shelf region along shiptrack 3.}
 
\label{fig:tape3}
 
\end{figure}

\begin{table*}

\begin{center}

\caption{Average Phi Values}

\label{tab:avg}

\begin{tabular}{ccccc}

\\[-.35cm]

\hline

\\[-.35cm]

Region & Grab Samples & Inversion & Sediment Type & \% Agreement\\

\hline

\\[-.35cm]

Slope & $1.92\pm0.36$ & $2.02\pm0.24$ & Medium--fine sand & $100\%$\\

Shelf & $0.82\pm0.71$ & $0.92\pm2.04$ & Medium--coarse sand & $86\%$\\

\hline

\\[-.35cm]

\end{tabular}

\end{center}

\end{table*}

\section{Conclusions}

The inversion results are in good agreement with the ground truth
(both in sediment type, i.e, sand, and in grade of sand) for the slope region where the sediment
layer is known to be relatively deep and homogeneous (near the
continental shelf break). The seafloor in
the shelf area, on the other hand, is known to have
little or no sediment layer and shells, rocks, etc., at the
bottom. The inversion from the shelf region also agrees in sediment
type with most grab samples, however, there is often a discrepancy in
grade of sand. Moreover, in a few cases the phi value from the inversion
is the lower limit (-1, i.e., gravel/coarse sand). Because the sonar frequency is 12 kHz, the sound will
penetrate any sediment layer less than about 13 cm (wavelength) deep and
interact with the hard subbottom. The backscattering strength
predicted by the APL-UW model in this case will be invalid.

Of the 129 inversions for the three shiptracks, 119 of them (92\%)
agree with the nearest grab sample in sediment type. Average values and standard
deviations are listed in Table~\ref{tab:avg} along with the percent
agreement of the inversion phi values with the nearest grab sample. 

We believe the inversion method described here is promising for
determining sediment type in areas of relatively homogeneous sediment and at least a few tens
of centimeters deep. This process currently also provides an approximation for thin sediment layers or sediment with heterogeneous mixtures. 

\section*{Acknowledgments}

This project was supported by the Space and Naval Warfare Systems
Command (SPAWAR).
The authors thank Mr. Brent Bartels (PSI) for help with the data
processing software and Dr. Fred Bowles (NRL) for ground truth
analysis. Useful discussions with Mr. Will Avera (NRL) are also acknowledged.

\nocite{*}
\bibliographystyle{IEEE}
%%%%%\bibliography{bib-file}  % commented if *.bbl file included, as
%%%%%see below

%%%%%%%%%%%%%%%%% BIBLIOGRAPHY IN THE LaTeX file !!!!! %%%%%%%%%%%%%%%%%%%%%%%%
%% This is nothing else than the IEEEsample.bbl file that you would         
%%
%% obtain with BibTeX: you do not need to send around the *.bbl file        
%%
%%---------------------------------------------------------------------------%%
%

\end{document}